\documentclass[12pt,a4paper,dvips]{article}
\usepackage{a4p}
\usepackage{cite,mcite}
\usepackage{graphicx}
\usepackage{physics}
\usepackage{l3_title,ifthen,Lep}
\usepackage[hang]{caption}
\journalname{Phys.~Lett.~B}
\date{November 18, 2003}
\preprint{2003-074}
\Lep{2}
\newlength{\capindent}
\newlength{\capwidth}
\newlength{\figwidth}
\newcommand{\icaption}[2][!*!,!]{\hspace*{\capindent}%
  \begin{minipage}{\capwidth}
    \ifthenelse{\equal{#1}{!*!,!}}%
      {\caption{#2}}%
      {\caption[#1]{#2}}
  \end{minipage}}
\newcommand{\rtsp}{\ensuremath{\sqrt{s^{\prime}}}}
\newcommand{\mz}{\ensuremath{m_{\rm Z}}}
\newcommand{\meff}{\ensuremath{m_{\rm eff}}}
\newcommand{\eer}{\ensuremath{\rm e^{+}e^{-}\rightarrow}}
\begin{document}
\begin{titlepage}
\title{Measurement of the Z-boson mass\\ 
       using \boldmath{${\rm e}^{+}{\rm e}^{-}\rightarrow\Zo\gamma$} events\\ 
       at centre-of-mass energies above the Z pole}

\author{The L3 Collaboration}

%
%
\begin{abstract}
Events from the $\rm e^{+}e^{-} \rightarrow Z\gamma$ process with hard initial-state radiation collected with the L3 detector
at centre-of-mass energies between 183 \GeV{} and 209 \GeV{} are used to measure the mass of the \Zo{} boson.
Decays of the Z boson into hadrons or muon pairs are considered and the Z mass is determined to be $91.272 \pm 0.032 \stat{} \pm 0.033~{\rm(syst.)} \GeV{}$,
in agreement with the value measured at the \Zo{} resonance.
Alternatively, assuming this measured value of the Z mass, the method determines the LEP centre-of-mass energy,
found to be $175\pm 68 \stat{} \pm 68~{\rm (syst.)} \MeV{}$ lower than the nominal value.
\end{abstract}

\submitted

\end{titlepage}

%
%

\section{Introduction}
At the LEP collider, operating at centre-of-mass energies, \rts{}, above the Z peak,
the process $\ee \rightarrow {\rm f}\bar{\rm f}\gamma$ frequently occurs with hard initial-state radiation (ISR).
Due to the strong resonance in the reaction $\ee \rightarrow \Zo \rightarrow {\rm f}\bar{\rm f}$,
the energy of the ISR photon is such that the mass of the fermion pair recoiling against the photon
is often close to the mass of the Z boson, \mz.
The strong forward peaking of the ISR process yields photons which usually remain undetected in the beampipe
but in a small fraction of events they are detected.

Data recorded with the L3 detector
\cite{L3DET} at $\sqrt{s} = 183 - 209 \GeV{}$ with a total integrated luminosity of 685~\pb{},
detailed in Table~\ref{tab_ebins}, are used to extract the mass of the Z boson, using Z decays into quark or muon pairs.
The mass of the hadronic system after applying a kinematic fit, \meff, is directly related to \mz.
In the case of muon pairs the event kinematics is fully determined by the measurement of the muon scattering angles,
making use of the excellent polar angle resolution of the muon spectrometer.
From the measured or reconstructed photon energy, $E_\gamma$, the effective centre-of-mass energy of the muon pair,
$\sqrt{s^{\prime}}$, is calculated as
\begin{equation}
  \rtsp{} = \rts{}~\sqrt{1-\frac{2\; E_{\gamma}}{\rts{}}}.
  \label{eqn_sprime}
\end{equation}
An unbinned likelihood fit is applied to the distribution of \meff{} or \rtsp{} to extract the Z mass.
The value is compared with the precision measurement of L3,
derived from the Z-lineshape scan at centre-of-mass energies around the Z pole~\cite{l3-202-acc}.
This comparison serves as a cross check of the W-mass measurement which uses similar techniques.
Assuming the Z mass measured around the Z pole, the method provides a measurement
of the LEP centre-of-mass energy which is compared to the one determined by the LEP energy working group~\cite{lep-energy-2000}.

%
%

\section{Monte Carlo simulation}
The Standard Model predictions for the different final states are determined with the following Monte Carlo programs:
KK2F~\cite{mc-kk2f} for $\rm \eer q\bar{q}(\gamma)$, $\rm\eer\mu^{+}\mu^{-}(\gamma)$ and $\rm \eer \tau^{+}\tau^{-}(\gamma)$,
BHWIDE~\cite{mc-bhwide} for $\rm \eer \ee(\gamma)$,
PHOJET~\cite{mc-phojet} for $\rm \eer \ee hadrons$,
DIAG36~\cite{mc-diag36} for $\rm \eer e^{+}e^{-}\mu^{+}\mu^{-}$,
KORALW~\cite{mc-koralw} for $\rm \eer W^{+}W^{-}$, with the exception of the $\rm e\nu q\bar{q}^{\prime}$ final state,
which is modelled by EXCALIBUR~\cite{EXCALIBUR},
and PYTHIA~\cite{mc-pythia} for $\eer \Zo{}\Zo{}$ and $\rm \eer Ze^{+}e^{-}$.
The response of the L3 detector is modelled with the GEANT \cite{GEANT-} detector simulation program
which includes the effects of energy loss, multiple scattering and showering in the detector material.
The GEISHA program~\cite{GEISHA} is used to simulate hadronic interactions in the detector.
Time-dependent detector inefficiencies are taken into account in the simulation.

%
%

\section{Selection of hadronic events}
\begin{figure}
  \begin{minipage}{.49\textwidth}
    \includegraphics[width=\textwidth]{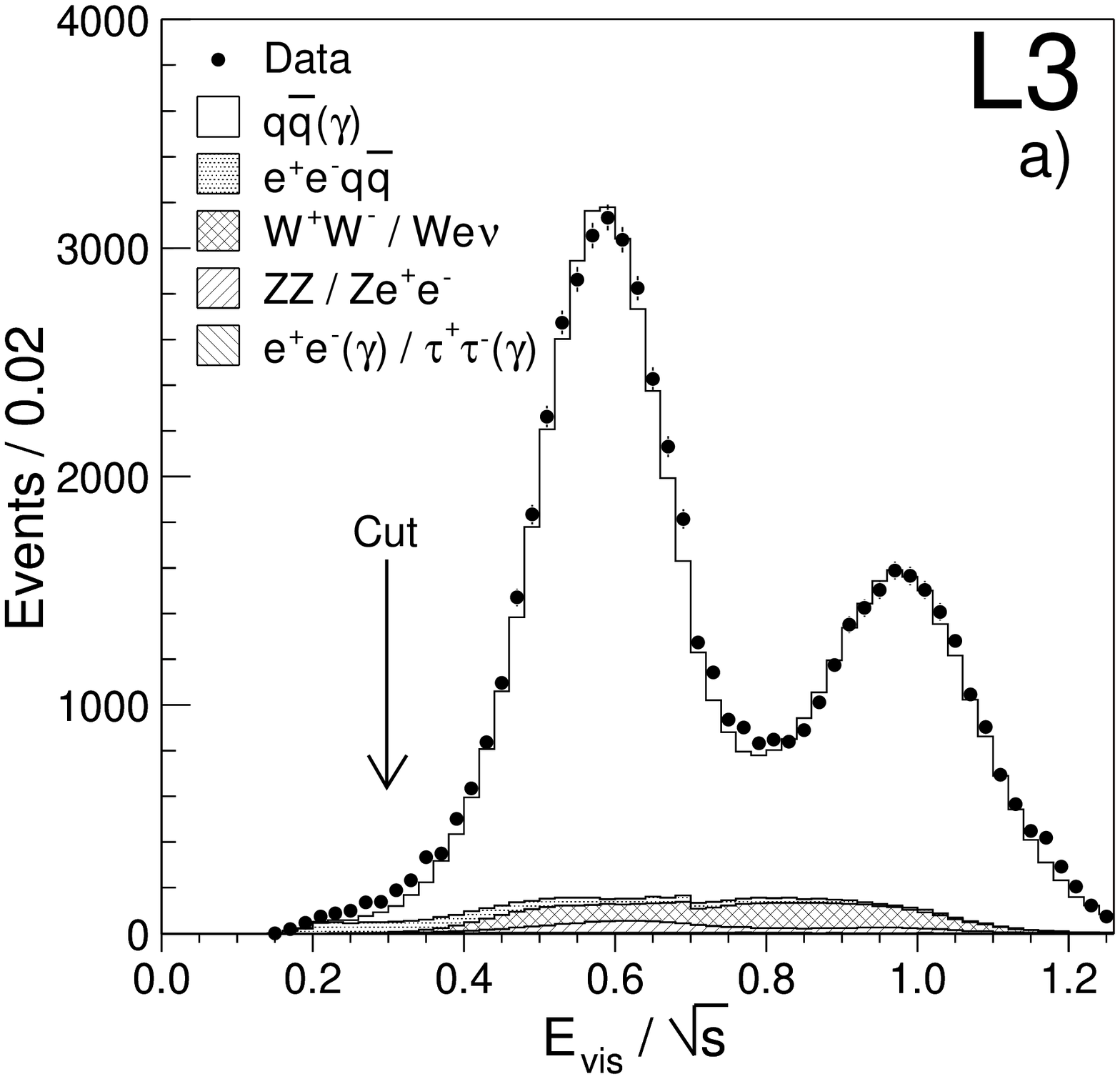}
  \end{minipage}
  \hfill
  \begin{minipage}{.49\textwidth}
    \includegraphics[width=\textwidth]{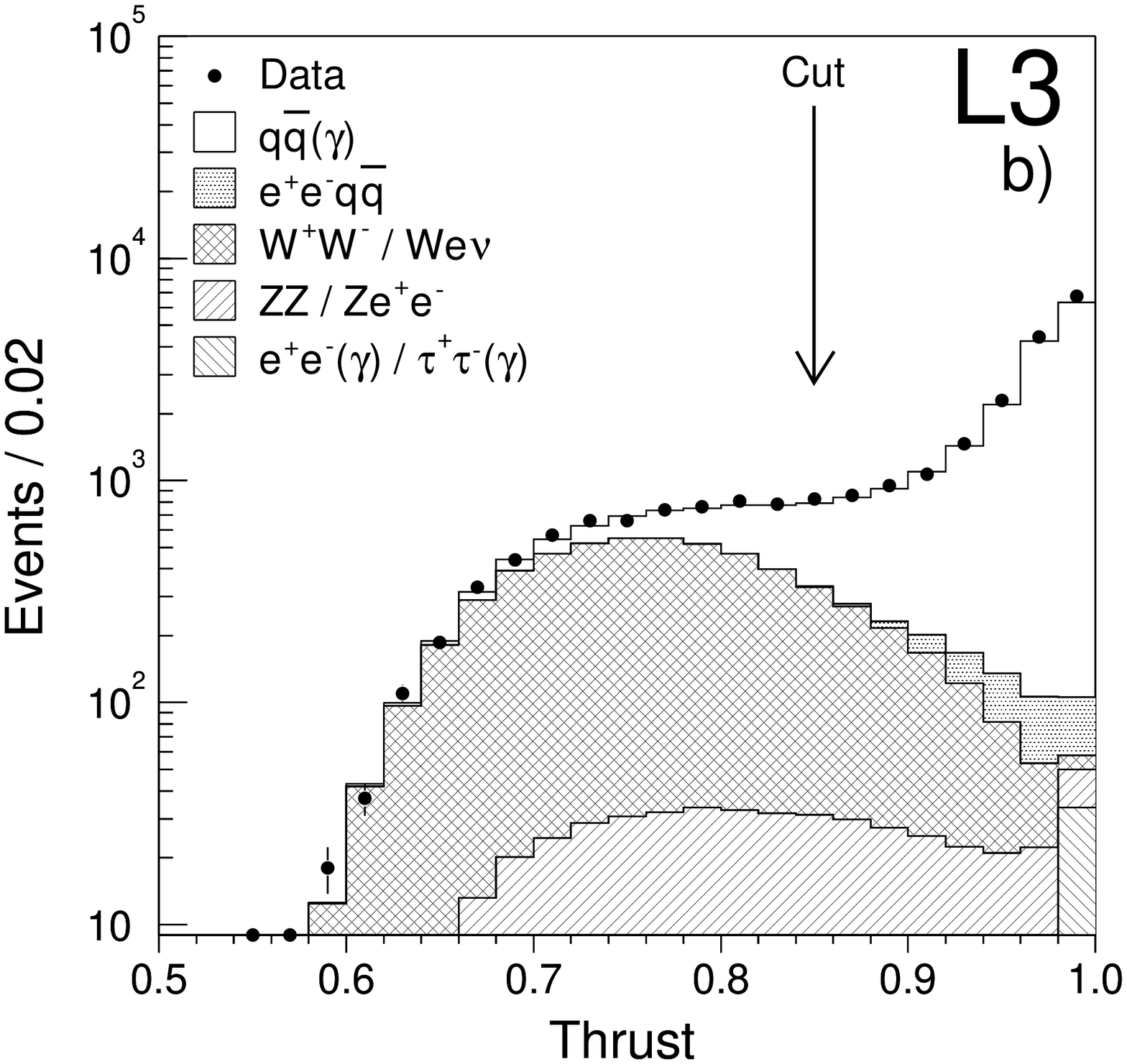}
  \end{minipage}

  \begin{minipage}{.49\textwidth}
    \includegraphics[width=\textwidth]{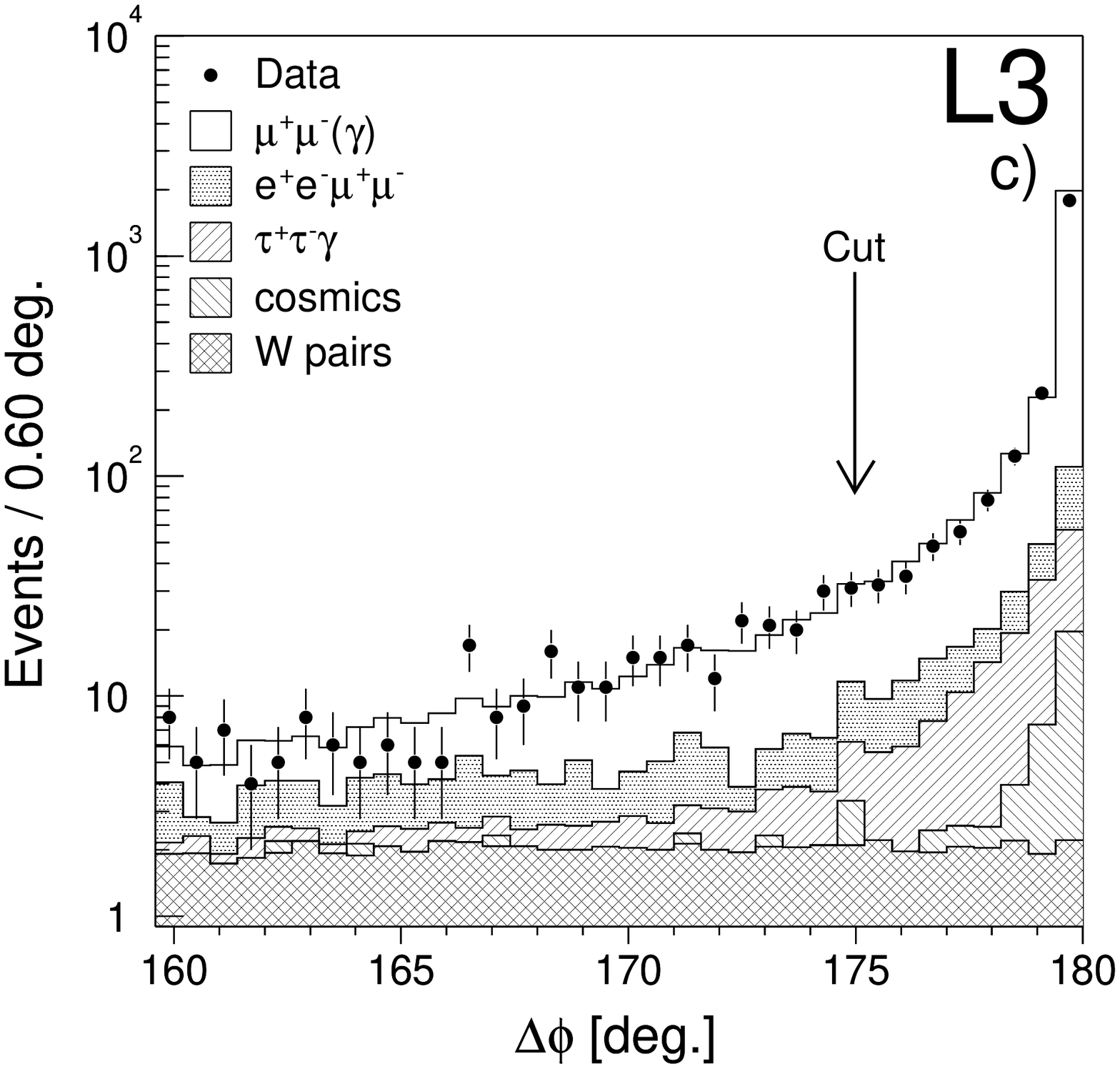}
  \end{minipage}
  \hfill
  \begin{minipage}{.49\textwidth}
    \includegraphics[width=\textwidth]{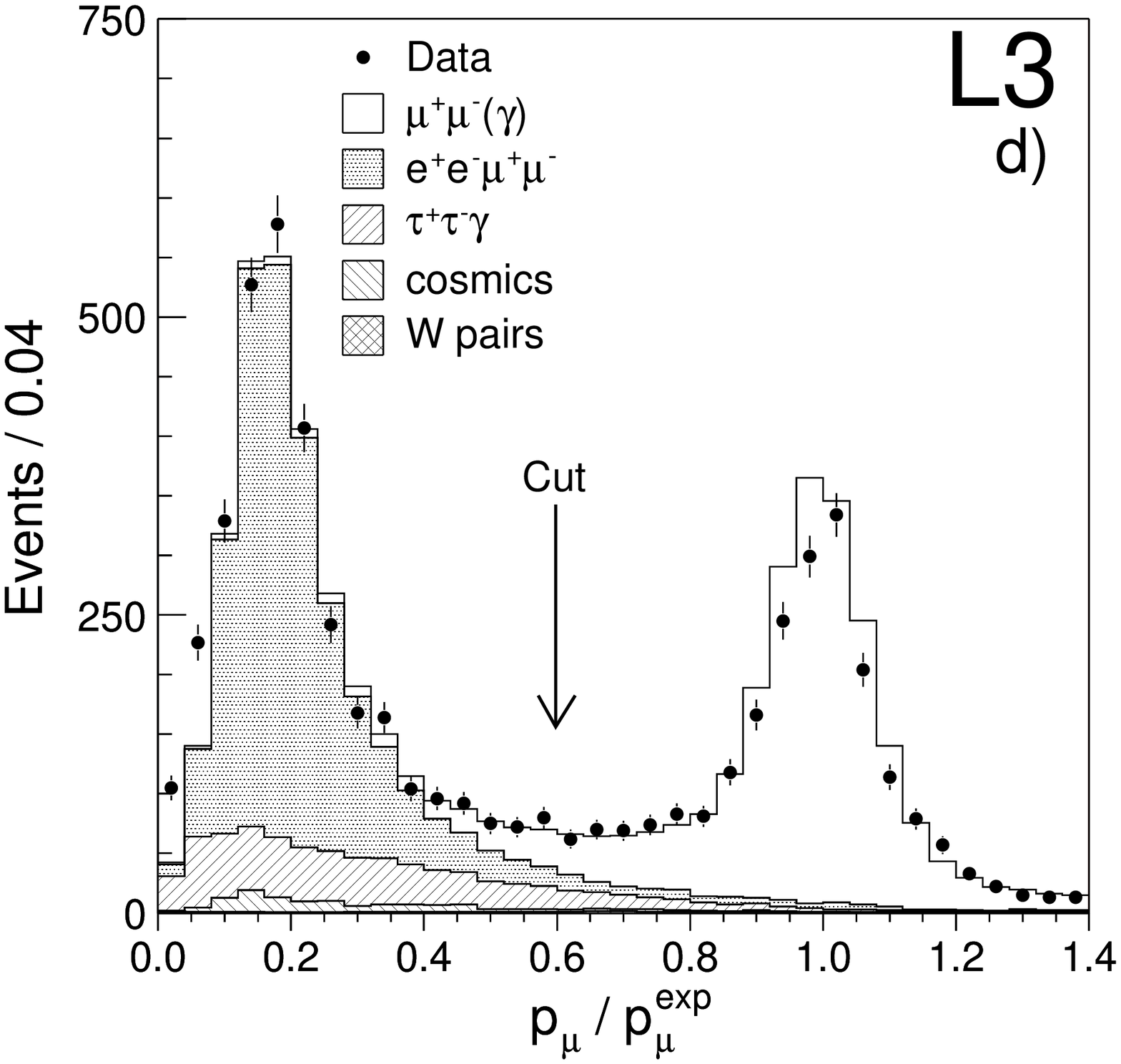}
  \end{minipage}
  \caption{
    Variables used for the event selections:
    a) visible energy, normalised to \rts,
    b) thrust in the centre-of-mass frame of the jets,
    c) angle between the muons in the plane perpendicular to the photon
    and d) measured muon momentum, normalised to the expected momentum.
  }
  \label{histo_qq}
\end{figure}

To remove purely leptonic final states the event must have more than 12 calorimetric clusters.
In addition, the transverse energy imbalance must be less than $25\%$ of the visible energy
and the sum of the cluster energies must be greater than $30\%$ of \rts, as shown in Figure~\ref{histo_qq}a.

Hadronic final states from two-photon collisions are typically boosted along the beam direction.
To reject these events we require the longitudinal energy imbalance not to exceed $80\%$ of the visible energy. 
Additionally, the sum of the transverse energies of the hadronic clusters has to be greater than $15\%$ of \rts.
Both cuts effectively remove the two-photon background.

Four-jet final states from \W{}- and \Zo{}-pair production have a total hadronic energy close to \rts.
The clusters in these events are distributed more spherically than in the case of two-jet events from $\Zo{}\gamma$ production.  
For events with a sum of the hadronic energy greater than $70\%$ of \rts,
the particles are boosted to the rest frame of the hadronic system.
A cut on the thrust, $T$, of the boosted event, $T > 0.85$, removes four-jet events as shown in Figure~\ref{histo_qq}b.

%
%

\section{Selection of muon-pair events}
Muons are reconstructed from tracks in the muon chambers in a fiducial volume of $\vert\cos{\theta}\vert < 0.9$.
If only one muon track is present in the event, the signature of an additional minimum-ionizing particle in the inner detectors is required.
Hadronic events are suppressed by requiring less than 15 calorimetric clusters.
To reject cosmic-ray background, the time measured by one scintillator matched in azimuthal angle to a muon track
has to be consistent with that of beam crossing within 5~ns.  
In addition, the distance of closest approach of at least one of the muon tracks to the interaction point
in the plane perpendicular to the beam must be less than 1~mm.  

The two muons are usually almost back-to-back in the plane perpendicular to the emitted photon.
Therefore, the angle between the muon directions in that plane, $\Delta\phi$, is required to be greater than $175^{\circ}$,
as shown in Figure~\ref{histo_qq}c.
The measured momentum, $p_{\mu}$, of the muon which has the largest polar angle, $\theta_{1}$, has to be greater than 60\%
of the corresponding expected momentum, $p_{\mu}^{\rm exp}$, calculated according to
\begin{equation}
  p_{\mu}^{\rm exp} = \sqrt{s}~
                      \frac{\sin{\theta_{2}}}{\sin{\theta_{1}} + \sin{\theta_{2}} + \vert\sin{(\theta_{1} + \theta_{2})}\vert },
  \label{eqn_pmurec}
\end{equation}
where $\theta_{2}$ is the polar angle of the other muon.
Figure~\ref{histo_qq}d shows the distribution of $p_{\mu}/p_{\mu}^{\rm exp}$.

%
%

\section{Mass reconstruction}
In each selected event we search for an isolated photon.
A photon candidate is a calorimetric cluster with an energy greater than 30\% of \rts.
It has to be isolated, with at most one additional cluster within $15^{\circ}$ of the candidate direction.
In addition, the angle between the photon and the nearest track must be greater than $4.6^{\circ}$.

\begin{figure}
  \includegraphics[width=.99\textwidth]{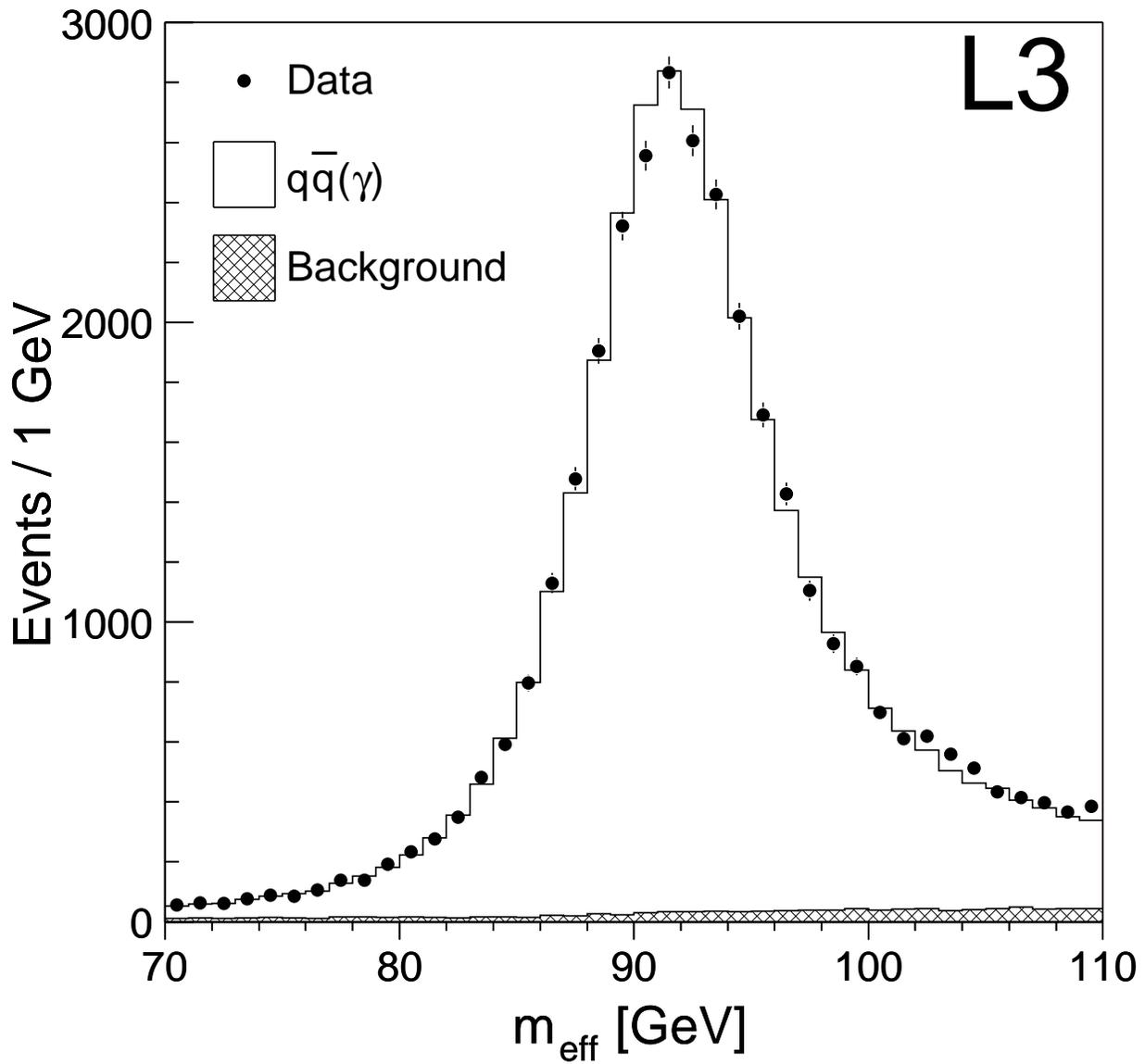}
  \caption{Distribution of the effective mass of the two-jet system after kinematic fit.
           The solid line shows the result of the fit which determines the Z boson mass.
  }
  \label{histo_minv}
\end{figure}

Hadronic events are forced into two jets using the DURHAM \cite{durham} algorithm, excluding photon candidates, if any.
The pion mass is assigned to energy depositions matched with a charged track
while clusters without a matching track are treated as massless.
A kinematic fit is applied to the event, where the measured quantities are varied
within their resolution and four-momentum conservation is imposed.
If no high-energy photon is observed in the detector, a single photon is assumed to have escaped undetected along the beam direction.
This improves the resolution of the two-jet effective mass by a factor of about 3.
The distribution of the fitted mass, \meff{}, is shown in Figure~\ref{histo_minv}.

\begin{figure}
  \begin{center}
    \includegraphics[width=.99\textwidth]{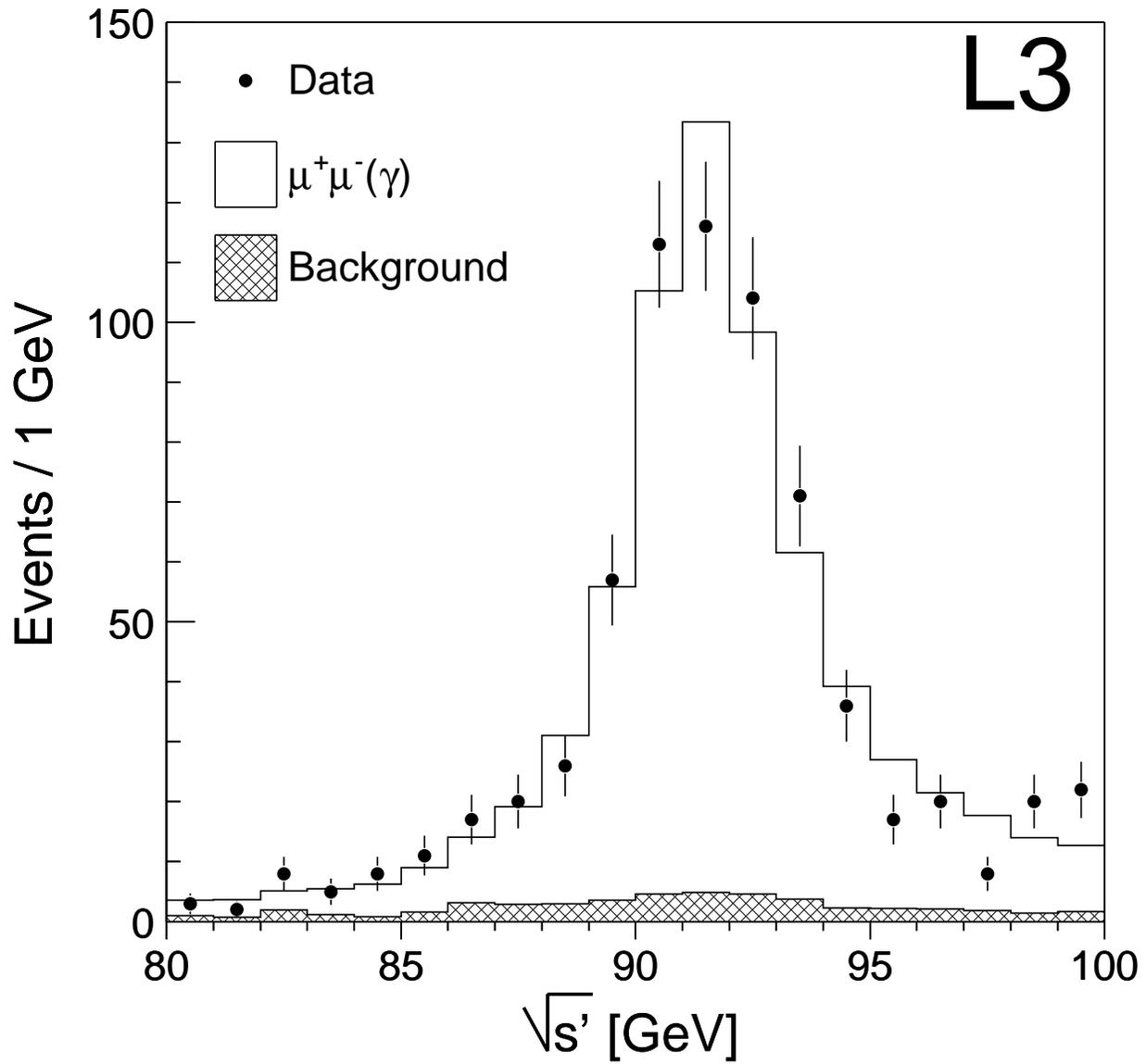}
    \caption{ 
      Distribution of the effective centre-of-mass energy for muon-pair events.
      The solid line shows the result of the fit which determines the Z boson mass.
    }
    \label{histo_muon_rtsp}
  \end{center}
\end{figure}

In muon-pair events, the photon energy is calculated from the reconstructed muon momenta using three-particle kinematics as
\begin{equation}
  E_{\gamma} =
  \sqrt{s}~\frac{\vert\sin(\theta_{1}+\theta_{2})\vert}
  {\sin\theta_{1}+\sin\theta_{2}+\vert\sin(\theta_{1}+\theta_{2})\vert}\;.
  \label{eqn_egamma}
\end{equation}
The $\theta_{i}$ are the angles between the muons and the photon direction.
If no  photon is detected they are the angles between the muons and the beam direction.
The effective centre-of-mass energy of the muon pair, \rtsp, is calculated using Formula~\ref{eqn_sprime}
and its distribution is shown in Figure~\ref{histo_muon_rtsp}.

%
%

\section{Fit method}
Events with hard ISR are selected by requiring
\[ 70\GeV{} < m_{\rm eff} < 110\GeV{} \]
for hadronic events and
\[ 80\GeV{} < \rtsp{} < 100\GeV{} \]
for muon-pair events.
Data samples of 34081 and 799 events are selected for hadrons and muon pairs, respectively,
with background levels of 3.2\% and 6.2\%.

To extract the Z mass from this data sample, an unbinned likelihood fit is applied to the measured differential cross section as a function of \meff{} or \rtsp.
The likelihood is defined as the product of the normalised differential cross sections of each event after selection cuts:
\begin{equation}
  L(m_{\Zo}^{\rm fit}) = \prod_{i} 
   \frac{\frac{\mathrm{d}\sigma}{\mathrm{d}\xi}(\xi_i,m_{\Zo}^{\rm fit}) +
         \frac{\mathrm{d}\sigma_{\mathrm{BG}}}{\mathrm{d}\xi}(\xi_i)}
        {\sigma(m_{\Zo}^{\rm fit}) + \sigma_{\mathrm{BG}}}\;,
  \label{eqn_possden}
\end{equation}
where $m_{\Zo}^{\rm fit}$ is the Z mass varied during the fit and $\xi$ represents
$m_{\rm eff}$ for the hadronic events and \rtsp{} for the muon-pair events.
The total and differential accepted cross sections of the signal are denoted as $\sigma$ and $\mathrm{d}\sigma/\mathrm{d}\xi$,
while $\sigma_{\rm BG}$ and $\mathrm{d}\sigma_{\rm BG}/\mathrm{d}\xi$ are the total and differential cross sections of the background.

The box method \cite{BOXMETHOD} is used to obtain the accepted differential cross section.
This method takes into account both detector resolution and selection efficiency effects.
The accepted differential cross section is determined by averaging Monte Carlo events
inside a $\xi$ bin centered around each data event.
The number of signal Monte Carlo events is scaled such that it agrees with the measured total cross section,
while the number of background events is normalised to the integrated luminosity.
This ansatz assigns fluctuations in the number of data events to the signal.
Stable results are obtained with bin sizes chosen such that 500 signal Monte Carlo events are contained in each \meff{} bin and 250 signal events in each \rtsp{} bin.
The size of the bins ranges from about 20~\MeV{} in the \meff{} peak to about 6~\GeV{} in the \meff{} tails
and from 50~\MeV{} in the \rtsp{} peak to 3~\GeV{} in the \rtsp{} tails.

To simulate the effect of different values of the Z-boson mass with a finite number of Monte Carlo events,
the signal Monte Carlo events are reweighted using the weights 
\begin{equation}
        \frac{\rm d\sigma}{{\rm d}\rtsp}
        \left(\rtsp_i^{\rm ~gen}, \mz^{\rm fit}\right) \Biggm/
        \frac{\rm d\sigma}{{\rm d}\rtsp}
        \left(\rtsp_i^{\rm ~gen}, \mz^{\rm MC}\right)\;,
\label{eqn_weights}
\end{equation}
where $\rtsp_{i}^{\rm ~gen}$ is the generated effective centre-of-mass energy of each event and
$\mz^{\rm MC}$ is the value of the \Zo{} mass that is used in the Monte Carlo generation.
The calculation of the differential cross section, ${\rm d\sigma/d}\rtsp$, up to $\mathcal{O}(\alpha^{2})$ of Reference \citen{Bardin:1989qr} is used.

To confirm the linearity of the fitting method as well as to test for
any possible bias, five Monte Carlo samples, each with about ten times
the statistics of the data, are generated with different \Zo{} masses
between 90~\GeV{} and 92~\GeV{}.  The correct values for the mass of
the Z boson are found by the fitting method within the statistical
precision of the test.

%
%

\section{\label{chap_syst}Systematic uncertainties}
The systematic uncertainties for the Z mass determination are listed in Table~\ref{tab_syst} and are detailed below.

To estimate the uncertainty due to hadronisation,
three different Monte Carlo samples are generated which have the same underlying two-fermion events,
but different hadronisation schemes, modelled by the programs ARIADNE~\cite{ARIADNE}, HERWIG~\cite{mc-herwig} and PYTHIA.
As the jets are formed from clusters that are either massless or assigned the pion mass,
a different kaon or baryon content in data and Monte Carlo would lead to a bias in the extracted Z mass.
For each of the three models the Monte Carlo events were reweighted in order to reproduce
the mean kaon and proton multiplicity measured on the Z peak \cite{PDB}.
The mass of the jets plays a role in the kinematic fit and
the differences of the jet masses produced by the different hadronisation schemes are also considered.
A reweighting method is applied to reproduce the jet masses measured in data.
The root-mean-square of the various Z masses obtained with the three Monte Carlo programs after the different reweighting
procedures is assigned as the hadronisation systematic uncertainty.

\begin{figure}
  \begin{minipage}{.49\textwidth}
    \includegraphics[width=\textwidth]{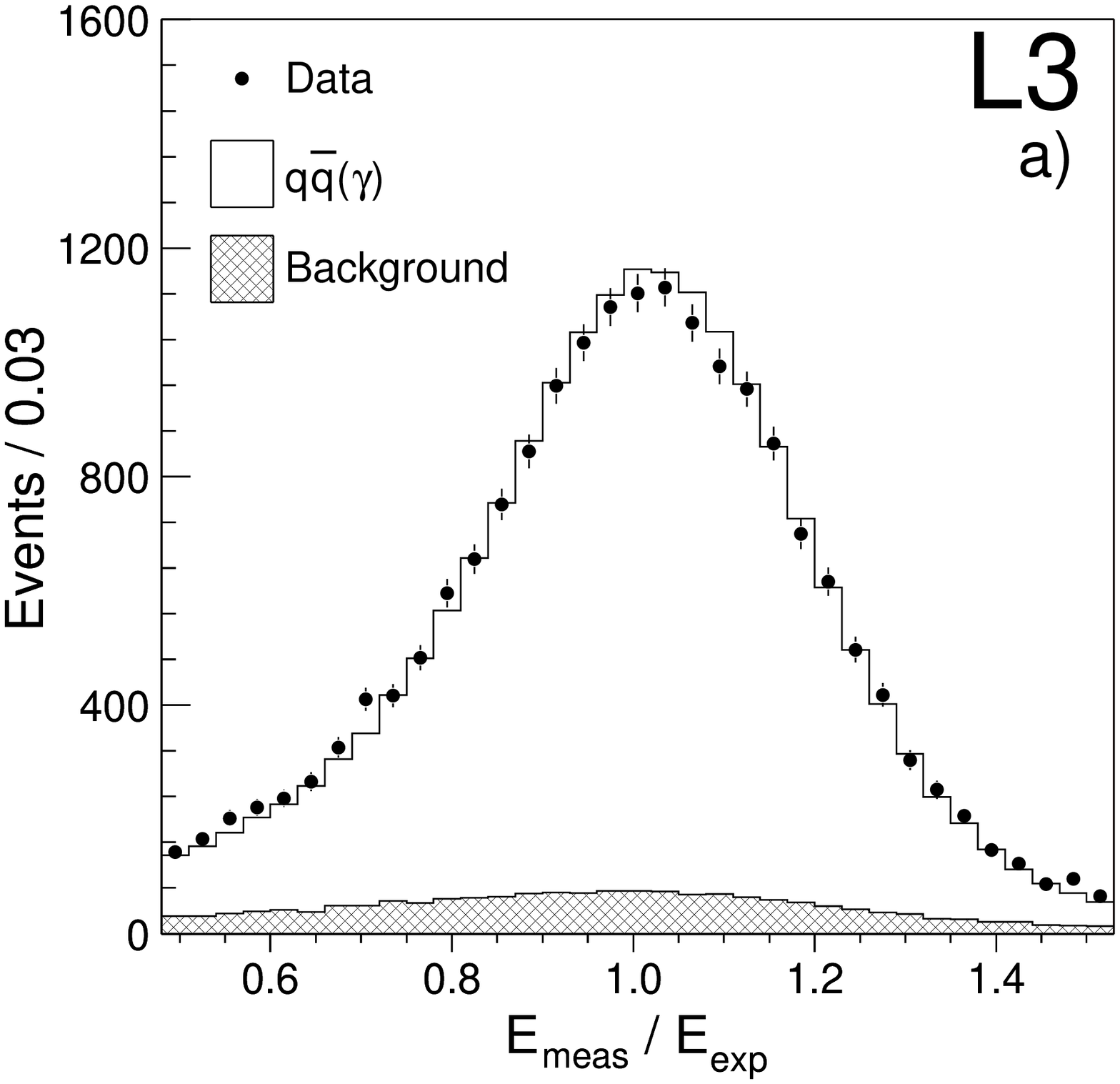}
  \end{minipage}
  \hfill
  \begin{minipage}{.49\textwidth}
    \includegraphics[width=\textwidth]{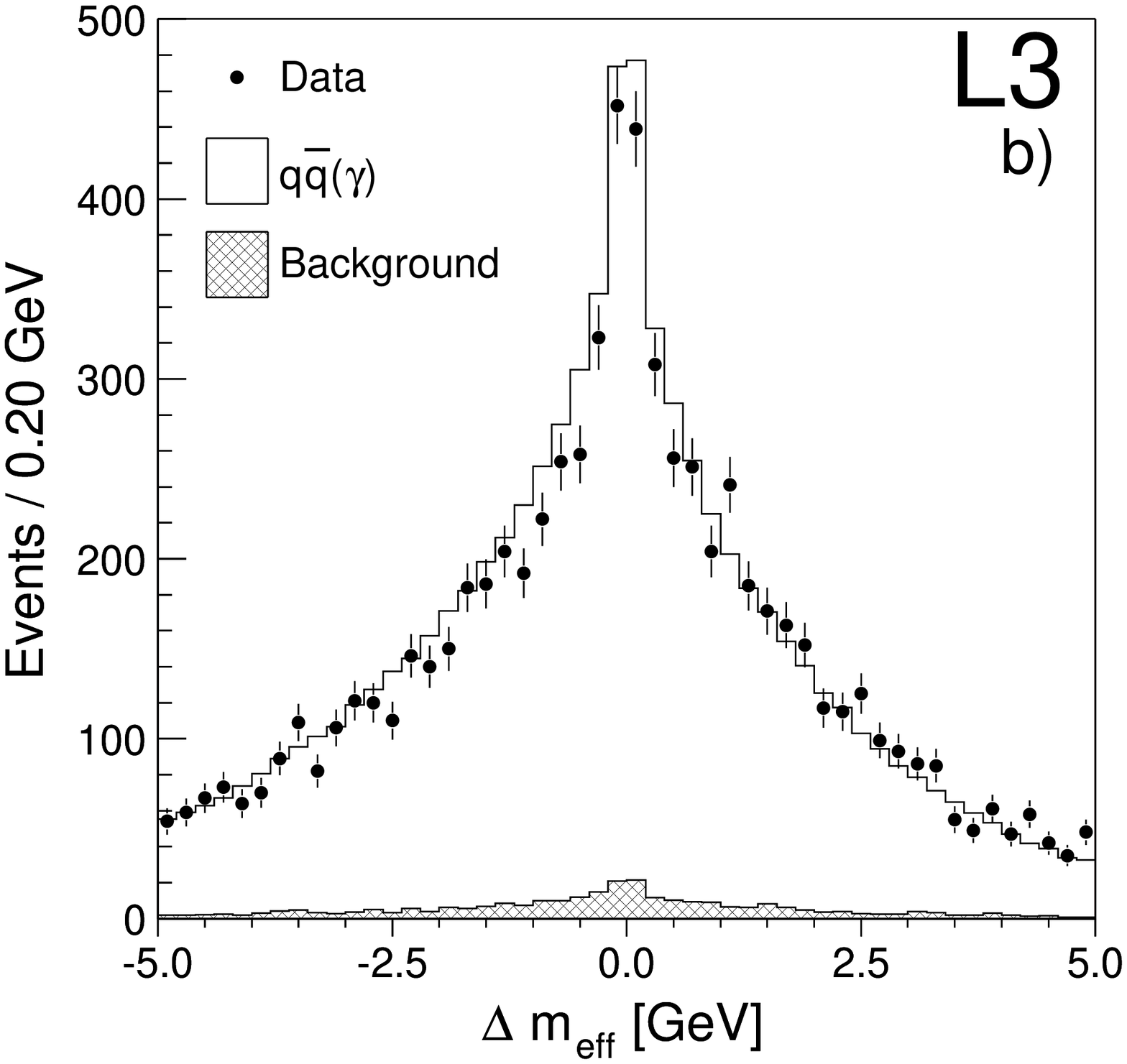}
  \end{minipage}

  \begin{minipage}{.49\textwidth}
    \includegraphics[width=\textwidth]{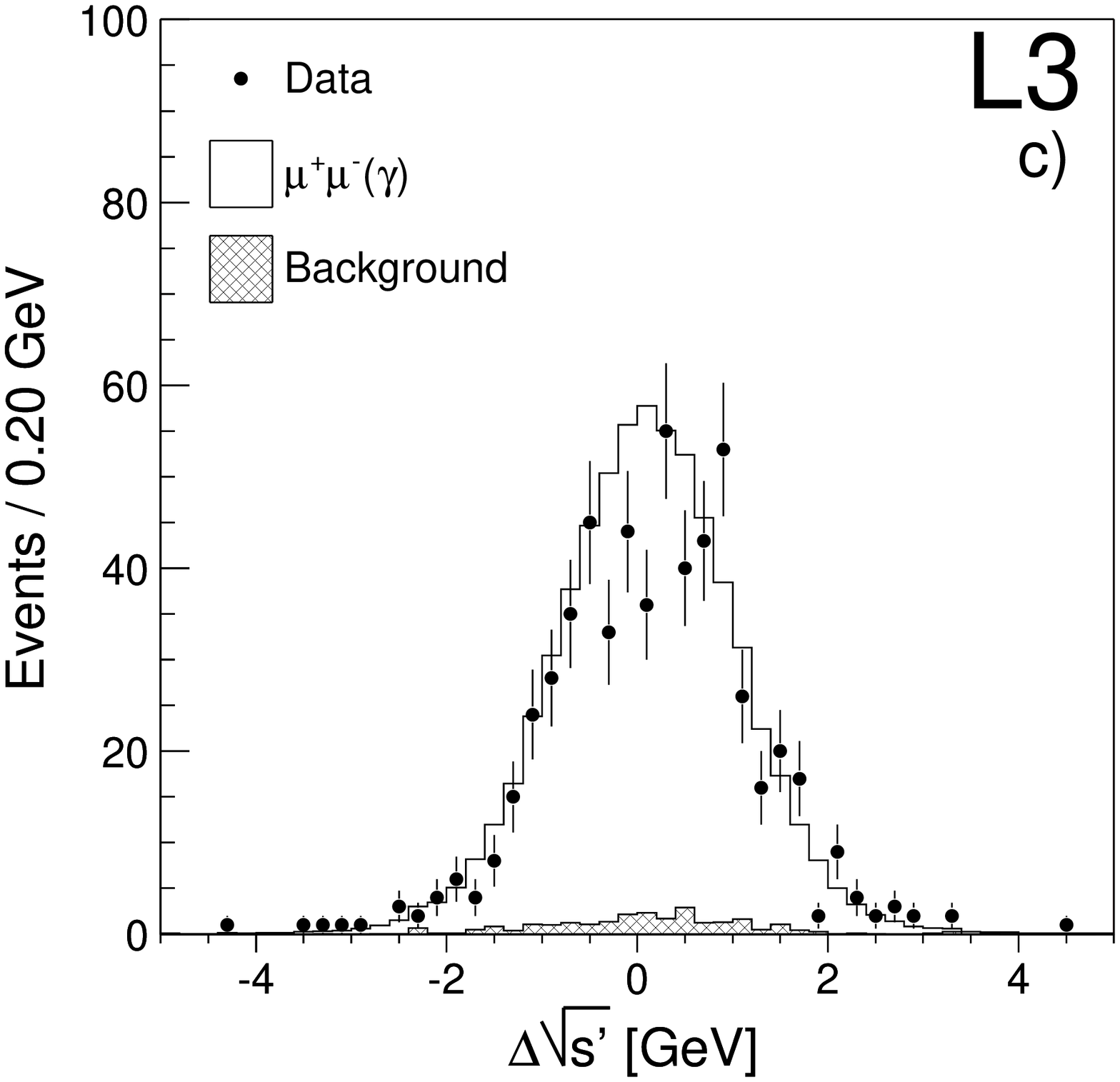}
  \end{minipage}
  \hfill
  \begin{minipage}{.49\textwidth}
    \includegraphics[width=\textwidth]{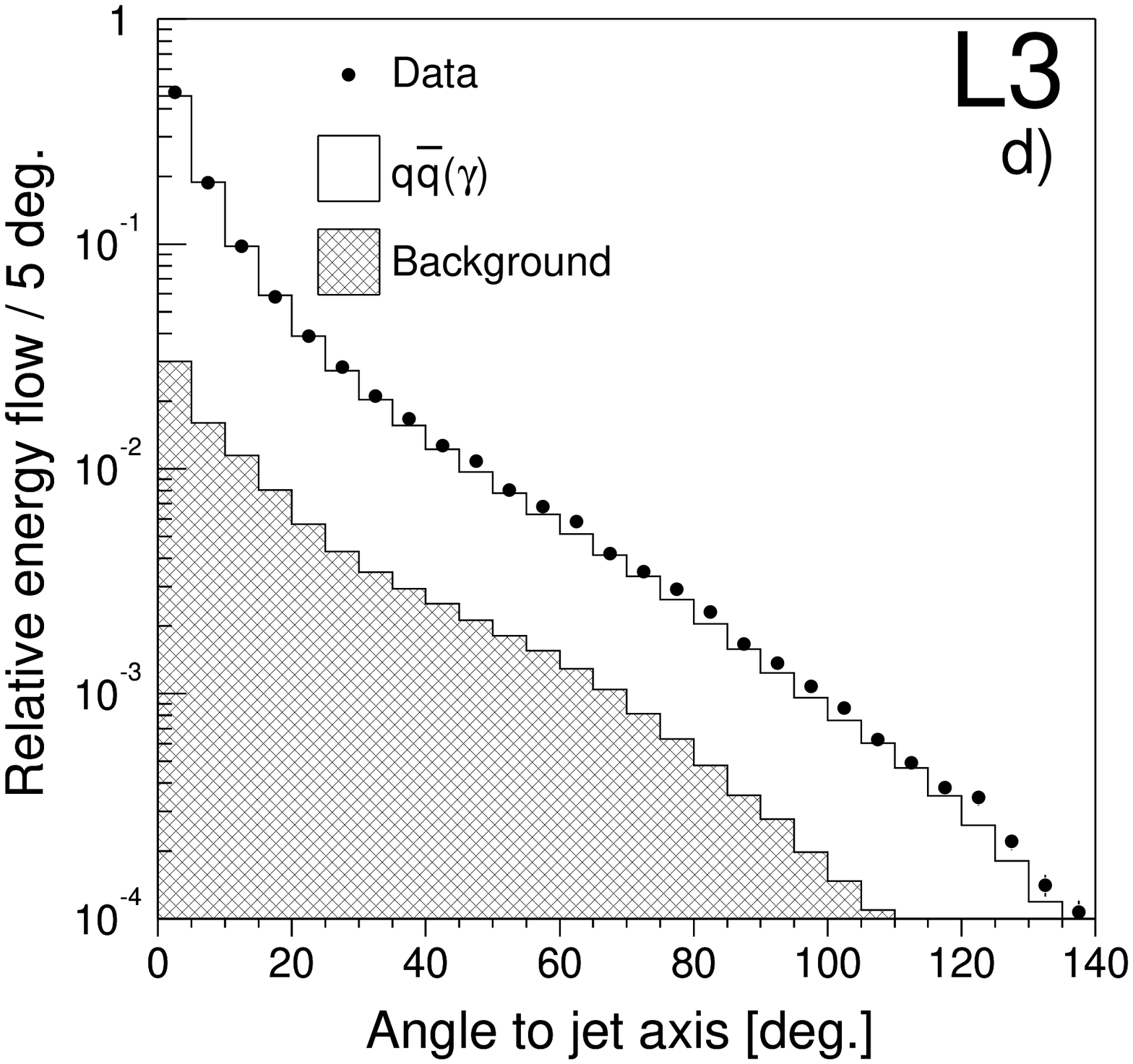}
  \end{minipage}
  \caption{
    Variables used for systematic studies:
    a) ratio of measured jet energy and jet energy calculated from jet angles,
    b) difference, per event, of the effective mass as obtained using cluster or track angles,
    c) difference, per event, of the effective centre-of-mass energy, obtained by using polar angles from the direction of the muons or the corresponding calorimetric clusters,
    d) relative energy flow in a jet as a function of the angular distance from jet axis.
  }
  \label{histo}
\end{figure}

The ratio of the measured jet energy to the energy calculated from the jet angles analogously to Formula~\ref{eqn_pmurec}, is shown in Figure~\ref{histo}a.
The jet energy scale is confirmed at the percent level.
The value of this ratio as a function of $\cos\theta$ allows for the recalibration of  the energy measurement of hadronic clusters.
Half of the difference to the mass obtained without recalibration is taken as systematic uncertainty.
In addition, we scale the energy of the individual energy depositions by $\pm$~0.5\%.
The difference to the value without scaling is taken as systematic uncertainty.
The total uncertainty from energy calibration is the sum in quadrature of the two contributions.

The uncertainty due to the measurement of the jet polar angle is estimated by repeating the analysis
with jets formed by calorimetric clusters that have a corresponding charged track and
jets formed by the same clusters but using the polar angles from the correlated charged track instead of the angle measured in the calorimeter.
The accuracy of the measurement of the muon polar angle is tested by comparing the default \rtsp{} spectrum
to the spectrum obtained using the angles of calorimetric clusters associated with the muons.
The differences of \meff{} or \rtsp{} are calculated on an event-by-event basis
and shown in Figures~\ref{histo}b and \ref{histo}c, respectively.
We assign half the difference of the average shifts of data and Monte Carlo as systematic uncertainties.

Additional clusters around detected photons,
wrongly assigned to one of the jets, would affect the reconstructed Z mass.
Removing all clusters in a $10^{\circ}$ cone around the photon direction results in a negligible mass shift.
In a similar way, calorimetric clusters from random noise, equally distributed in the calorimeters,
would affect the reconstructed mass if they are not described by the Monte Carlo.
The angular distribution of calorimetric energy relative to the corresponding jet axis is plotted in Figure~\ref{histo}d.
Good agreement between data and Monte Carlo is seen.
Removing all clusters outside a cone of $60^{\circ}$ half-opening angle around the jet axis yields a negligible mass shift.

For both hadronic events and muon-pair events the uncertainty in the background level is evaluated
by scaling the total cross section of the background Monte Carlo samples by $\pm 5\%$.

The theoretical uncertainty on the ISR modelling is tested by comparing results obtained with different ISR modelling schemes to different orders of $\alpha$.
This is done by using event weights given by the Monte Carlo generator KK2F as described in Reference~\citen{mc-kk2f}.
The differences between the results from $\mathcal{O}(\alpha^{2})$ calculations with Coherent Exclusive Exponentiation with and without 
ISR/FSR interference is assigned as systematic uncertainty.

The stability of the box method is tested by changing the box sizes used in the fit.
A small shift is observed and quoted as systematic uncertainty.
To test the results obtained with the maximum likelihood fit, we also apply a $\chi^{2}$ fit of the Monte Carlo histogram
to the measured \meff{} distribution by reweighting individual Monte Carlo events.
As a second test, we apply the maximum likelihood fit to the \rtsp{} spectrum of the hadronic event sample,
calculated using Formula \ref{eqn_egamma} modified to include the effect of non zero jet masses.
The results from both cross checks show no significant deviation from the default method.
The uncertainty from limited Monte Carlo statistics is also considered.

The LEP beam energy has an uncertainty of 10 to 20 \MeV{}~\cite{lep-energy-2000} depending on \rts.
The relative error on the Z-boson mass is the same as that on the beam energy.

The systematic uncertainties are treated as fully correlated between the different energy points,
except the uncertainty resulting from limited Monte Carlo statistics, which is treated as uncorrelated,
and the uncertainties on the beam energy, where the correlation matrix from Reference~\cite{lep-energy-2000} is used.
A total systematic uncertainty of 39~\MeV{} for the determination of the Z-mass from the hadronic channel and 21~\MeV{} from the muon channel is found.

%
%

\section{Results}
The results for different values of \rts{} are shown in Table \ref{tab_MZ_LL_KF}.
They are combined taking the systematic uncertainties into account.
The combination yields
\[ m_{\Zo}^{\rm qq} = 91.271 \pm 0.031\stat \pm 0.039~{\rm(syst.)} \GeV{} \]
for hadronic events and 
\[ m_{\Zo}^{\mu\mu} = 91.276 \pm 0.105\stat \pm 0.021~{\rm(syst.)} \GeV{} \]
for muon-pair events.
The results of the fit are shown in Figures \ref{histo_minv} and \ref{histo_muon_rtsp}.
Averaging the results obtained from the hadronic and muon pair samples, including all correlations, yields
\[ m_{\Zo}^{\rm meas} = 91.272 \pm 0.032\stat \pm 0.033~{\rm(syst.)} \GeV{}. \]
For the combination of hadron and muon-pair events, the uncertainty due to ISR modelling is treated as fully correlated.
For the beam energy uncertainties the correlation matrix of Reference~\citen{lep-energy-2000} is used.
This value is in good agreement with the precision mass measurement, $m_{\Zo{}} = 91.1898 \pm 0.0031 \GeV{}$~\cite{l3-202-acc}.

The measurement can also be interpreted as a determination of the LEP centre-of-mass energy, $\sqrt{s}^{\rm meas}$.
A difference between the measured mass, $m_{\Zo}^{\rm meas}$, and the precision mass, $m_{\Zo{}}$,
can be attributed to a deviation from the nominal value, $\sqrt{s}$:
\begin{equation}
  \Delta \sqrt{s} =
  \sqrt{s}^{\rm meas} - \sqrt{s} = -\sqrt{s} \frac{m_{\Zo{}}^{\rm meas}-m_{\Zo{}}}{m_{\Zo{}}}.
\end{equation}
The value obtained from the observed \Zo{} mass,
\[ \Delta \sqrt{s} = - 0.175 \pm 0.068 \stat\pm 0.068\sys{} \GeV{}, \]
is consistent with no shift. 

In conclusion, the Z mass measured in radiative events is in agreement with the determination at the Z pole,
validating the method used for the measurement of the mass of the W boson.
Interpreted as a determination of the centre-of-mass energy it agrees with the measurements by the LEP energy working group~\cite{lep-energy-2000}.

%
%

\newpage

\bibliographystyle{/home/achdsrv1/institut_3a/rosenbl/tex/biblio/l3stylem}

%
%

\newpage
\section*{Author List}
\typeout{   }     
\typeout{Using author list for paper 274 -  }
\typeout{$Modified: Jul 15 2001 by smele $}
\typeout{!!!!  This should only be used with document option a4p!!!!}
\typeout{   }
%
%
%
%
%
%

\newcount\tutecount  \tutecount=0
\def\tutenum#1{\global\advance\tutecount by 1 \xdef#1{\the\tutecount}}
\def\tute#1{$^{#1}$}
\tutenum\aachen            
\tutenum\nikhef            
\tutenum\mich              
\tutenum\lapp              
\tutenum\basel             
\tutenum\lsu               
\tutenum\beijing           
\tutenum\bologna           
\tutenum\tata              
\tutenum\ne                
\tutenum\bucharest         
\tutenum\budapest          
\tutenum\mit               
\tutenum\panjab            
\tutenum\debrecen          
\tutenum\dublin            
\tutenum\florence          
\tutenum\cern              
\tutenum\wl                
\tutenum\geneva            
\tutenum\hefei             
\tutenum\lausanne          
\tutenum\lyon              
\tutenum\madrid            
\tutenum\florida           
\tutenum\milan             
\tutenum\moscow            
\tutenum\naples            
\tutenum\cyprus            
\tutenum\nymegen           
\tutenum\caltech           
\tutenum\perugia           
\tutenum\peters            
\tutenum\cmu               
\tutenum\potenza           
\tutenum\prince            
\tutenum\riverside         
\tutenum\rome              
\tutenum\salerno           
\tutenum\ucsd              
\tutenum\sofia             
\tutenum\korea             
\tutenum\purdue            
\tutenum\psinst            
\tutenum\zeuthen           
\tutenum\eth               
\tutenum\hamburg           
\tutenum\taiwan            
\tutenum\tsinghua          

{
\parskip=0pt
\noindent
{\bf The L3 Collaboration:}
\ifx\selectfont\undefined
 \baselineskip=10.8pt
 \baselineskip\baselinestretch\baselineskip
 \normalbaselineskip\baselineskip
 \ixpt
\else
 \fontsize{9}{10.8pt}\selectfont
\fi
\medskip
\tolerance=10000
\hbadness=5000
\raggedright
\hsize=162truemm\hoffset=0mm
\def\r{\rlap,}
\noindent

P.Achard\r\tute\geneva\ 
O.Adriani\r\tute{\florence}\ 
M.Aguilar-Benitez\r\tute\madrid\ 
J.Alcaraz\r\tute{\madrid}\ 
G.Alemanni\r\tute\lausanne\
J.Allaby\r\tute\cern\
A.Aloisio\r\tute\naples\ 
M.G.Alviggi\r\tute\naples\
H.Anderhub\r\tute\eth\ 
V.P.Andreev\r\tute{\lsu,\peters}\
% ...
\newpage
\begin{list}{A}{\itemsep=0pt plus 0pt minus 0pt\parsep=0pt plus 0pt minus 0pt
                \topsep=0pt plus 0pt minus 0pt}
\item[\aachen]
 III. Physikalisches Institut, RWTH, D-52056 Aachen, Germany$^{\S}$
\item[\nikhef] National Institute for High Energy Physics, NIKHEF, 
     and University of Amsterdam, NL-1009 DB Amsterdam, The Netherlands
\item[\mich] University of Michigan, Ann Arbor, MI 48109, USA
\item[\lapp] Laboratoire d'Annecy-le-Vieux de Physique des Particules, 
     LAPP,IN2P3-CNRS, BP 110, F-74941 Annecy-le-Vieux CEDEX, France
\item[\basel] Institute of Physics, University of Basel, CH-4056 Basel,
     Switzerland
\item[\lsu] Louisiana State University, Baton Rouge, LA 70803, USA
\item[\beijing] Institute of High Energy Physics, IHEP, 
  100039 Beijing, China$^{\triangle}$ 
\item[\bologna] University of Bologna and INFN-Sezione di Bologna, 
     I-40126 Bologna, Italy
\item[\tata] Tata Institute of Fundamental Research, Mumbai (Bombay) 400 005, India
\item[\ne] Northeastern University, Boston, MA 02115, USA
\item[\bucharest] Institute of Atomic Physics and University of Bucharest,
     R-76900 Bucharest, Romania
\item[\budapest] Central Research Institute for Physics of the 
     Hungarian Academy of Sciences, H-1525 Budapest 114, Hungary$^{\ddag}$
\item[\mit] Massachusetts Institute of Technology, Cambridge, MA 02139, USA
\item[\panjab] Panjab University, Chandigarh 160 014, India.
\item[\debrecen] KLTE-ATOMKI, H-4010 Debrecen, Hungary$^\P$
\item[\dublin] Department of Experimental Physics,
  University College Dublin, Belfield, Dublin 4, Ireland
\item[\florence] INFN Sezione di Firenze and University of Florence, 
     I-50125 Florence, Italy
\item[\cern] European Laboratory for Particle Physics, CERN, 
     CH-1211 Geneva 23, Switzerland
\item[\wl] World Laboratory, FBLJA  Project, CH-1211 Geneva 23, Switzerland
\item[\geneva] University of Geneva, CH-1211 Geneva 4, Switzerland
\item[\hefei] Chinese University of Science and Technology, USTC,
      Hefei, Anhui 230 029, China$^{\triangle}$
\item[\lausanne] University of Lausanne, CH-1015 Lausanne, Switzerland
\item[\lyon] Institut de Physique Nucl\'eaire de Lyon, 
     IN2P3-CNRS,Universit\'e Claude Bernard, 
     F-69622 Villeurbanne, France
\item[\madrid] Centro de Investigaciones Energ{\'e}ticas, 
     Medioambientales y Tecnol\'ogicas, CIEMAT, E-28040 Madrid,
     Spain${\flat}$ 
\item[\florida] Florida Institute of Technology, Melbourne, FL 32901, USA
\item[\milan] INFN-Sezione di Milano, I-20133 Milan, Italy
\item[\moscow] Institute of Theoretical and Experimental Physics, ITEP, 
     Moscow, Russia
\item[\naples] INFN-Sezione di Napoli and University of Naples, 
     I-80125 Naples, Italy
\item[\cyprus] Department of Physics, University of Cyprus,
     Nicosia, Cyprus
\item[\nymegen] University of Nijmegen and NIKHEF, 
     NL-6525 ED Nijmegen, The Netherlands
\item[\caltech] California Institute of Technology, Pasadena, CA 91125, USA
\item[\perugia] INFN-Sezione di Perugia and Universit\`a Degli 
     Studi di Perugia, I-06100 Perugia, Italy   
\item[\peters] Nuclear Physics Institute, St. Petersburg, Russia
\item[\cmu] Carnegie Mellon University, Pittsburgh, PA 15213, USA
\item[\potenza] INFN-Sezione di Napoli and University of Potenza, 
     I-85100 Potenza, Italy
\item[\prince] Princeton University, Princeton, NJ 08544, USA
\item[\riverside] University of Californa, Riverside, CA 92521, USA
\item[\rome] INFN-Sezione di Roma and University of Rome, ``La Sapienza",
     I-00185 Rome, Italy
\item[\salerno] University and INFN, Salerno, I-84100 Salerno, Italy
\item[\ucsd] University of California, San Diego, CA 92093, USA
\item[\sofia] Bulgarian Academy of Sciences, Central Lab.~of 
     Mechatronics and Instrumentation, BU-1113 Sofia, Bulgaria
\item[\korea]  The Center for High Energy Physics, 
     Kyungpook National University, 702-701 Taegu, Republic of Korea
\item[\purdue] Purdue University, West Lafayette, IN 47907, USA
\item[\psinst] Paul Scherrer Institut, PSI, CH-5232 Villigen, Switzerland
\item[\zeuthen] DESY, D-15738 Zeuthen, Germany
\item[\eth] Eidgen\"ossische Technische Hochschule, ETH Z\"urich,
     CH-8093 Z\"urich, Switzerland
\item[\hamburg] University of Hamburg, D-22761 Hamburg, Germany
\item[\taiwan] National Central University, Chung-Li, Taiwan, China
\item[\tsinghua] Department of Physics, National Tsing Hua University,
      Taiwan, China
\item[\S]  Supported by the German Bundesministerium 
        f\"ur Bildung, Wissenschaft, Forschung und Technologie
\item[\ddag] Supported by the Hungarian OTKA fund under contract
numbers T019181, F023259 and T037350.
\item[\P] Also supported by the Hungarian OTKA fund under contract
  number T026178.
\item[$\flat$] Supported also by the Comisi\'on Interministerial de Ciencia y 
        Tecnolog{\'\i}a.
\item[$\sharp$] Also supported by CONICET and Universidad Nacional de La Plata,
        CC 67, 1900 La Plata, Argentina.
\item[$\triangle$] Supported by the National Natural Science
  Foundation of China.
\end{list}
}
\vfill


%
%

\newpage

\begin{table}
  \begin{center}
    \begin{tabular}{|c|c|c|c|}
      \hline
      Year     & $\left<\sqrt{s}\right>$ [\GeV{}] & Luminosity [$\rm pb^{-1}$]\\
      \hline
      1997     & 182.7                 & \phantom{0}55.5\\
      \hline
      1998     & 188.6                 & 176.8\\
      \hline
      1999     & 191.6                 & \phantom{0}29.8\\
      1999     & 195.5                 & \phantom{0}84.1\\
      1999     & 199.5                 & \phantom{0}83.3\\
      1999     & 201.8                 & \phantom{0}37.2\\
      \hline
      2000     & 204.8                 & \phantom{0}79.0\\
      2000     & 206.5                 & 130.8\\
      2000     & 208.0                 & \phantom{00}8.3\\
      \hline
      Total    &                       & 684.8\\
      \hline
    \end{tabular}
  \end{center}
  \caption{
    Centre-of-mass energies and corresponding integrated luminosities.
  }
  \label{tab_ebins}
\end{table}

\begin{table}
  \begin{center}
    \begin{tabular}{|l|c|c|}
	\hline
	 & Uncertainty [\MeV{}] & Uncertainty [\MeV{}]\\
        Source & (hadron channel)     & (muon channel)\\
	\hline
	Hadronisation           & 22           & $-$                    \\
	Energy calibration      & 16           & $-$                    \\
	Angular measurement     & 24           & \phantom{<}11          \\
	Background              & \phantom{0}3 & $<$1                   \\
	Initial-state radiation & \phantom{0}4 & \phantom{<}11          \\
	Box Size                & \phantom{0}2 & \phantom{<}\phantom{0}2\\
	Monte Carlo statistics  & \phantom{0}8 & \phantom{<}\phantom{0}9\\
	LEP energy              & 11           & \phantom{<}11          \\
	\hline
	Total                   & 39           & \phantom{<}21          \\
	\hline
    \end{tabular}
  \end{center}
  \caption{
    Systematic uncertainties on the Z boson mass.
  }
  \label{tab_syst}
\end{table}

\begin{table}
  \begin{center}
    \begin{tabular}{|c|c|c|}
      \hline
      & $m_{\Zo}^{\rm meas}$ [\GeV{}]      & $m_{\Zo}^{\rm meas}$ [\GeV{}]           \\
      $\left<\sqrt{s}\right>$ [\GeV{}] & (hadron channel) & (muon channel)   \\
      \hline
      $182.7$  & $91.286 \pm 0.095 \pm 0.046$ & $91.057 \pm 0.317 \pm 0.029$ \\
      $188.6$  & $91.290 \pm 0.057 \pm 0.042$ & $91.224 \pm 0.189 \pm 0.029$ \\
      $191.6$  & $91.402 \pm 0.143 \pm 0.046$ & $92.065 \pm 0.635 \pm 0.029$ \\
      $195.5$  & $91.467 \pm 0.089 \pm 0.046$ & $91.219 \pm 0.332 \pm 0.029$ \\
      $199.5$  & $91.144 \pm 0.094 \pm 0.046$ & $91.183 \pm 0.422 \pm 0.029$ \\
      $201.8$  & $91.073 \pm 0.142 \pm 0.046$ & $91.464 \pm 0.402 \pm 0.029$ \\
      $204.8$  & $91.369 \pm 0.101 \pm 0.042$ & $91.358 \pm 0.260 \pm 0.032$ \\
      $206.5$  & $91.107 \pm 0.081 \pm 0.042$ & $91.439 \pm 0.273 \pm 0.033$ \\
      $208.0$  & $91.329 \pm 0.331 \pm 0.042$ & $90.439 \pm 0.667 \pm 0.033$ \\
      \hline				   
      Combined & $91.271 \pm 0.031 \pm 0.039$ & $91.276 \pm 0.105 \pm 0.021$ \\
      \hline
    \end{tabular}
  \end{center}
  \caption{
    Results of the maximum likelihood fit for each \rts{} value.
    The first uncertainty is statistical, the second systematic.
  }
  \label{tab_MZ_LL_KF}
\end{table}

%
%

\newpage

\newpage

\end{document}